\documentclass[12pt]{article}
\usepackage{epsfig}

\def\beq{\begin{equation}}
\def\eeq#1{\label{#1}\end{equation}}
\def\eeqn{\end{equation}}

\def\beqa{\begin{eqnarray}}
\def\eeqa#1{\label{#1}\end{eqnarray}}
\def\eeqan{\end{eqnarray}}

\let\bar=\overbar

\def\Dslash{\not{\hbox{\kern-4pt $D$}}}
\def\dslash{\not{\hbox{\kern-2pt $\del$}}}

\def\msb{{\bar{\ssstyle M \kern -1pt S}}}

\def\Title#1{\begin{center} {\Large {\bf #1} } \end{center}}

\begin{document}

\Title{Electroweak di-boson production in ATLAS}

\bigskip\bigskip

%+\addtocontents{toc}{{\it J. Ebke}}
%+\label{EbkeStart}

\begin{raggedright}  

{\it Johannes Ebke\index{Ebke, J.}, on behalf of the ATLAS Collaboration\\
Fakult\"at f\"ur Physik, Ludwig-Maximilians-Universit\"at M\"unchen \\
Am Coulombwall 1, D-85748 Garching, Germany \\
johannes.ebke@physik.uni-muenchen.de}
\bigskip\bigskip
\end{raggedright}

\section{Introduction}

In the LHC era, it is crucial to gain a good understanding of the electroweak sector of the Standard Model (SM). Determining the cross-sections of electroweak processes is therefore of great importance, both as prerequisites to Higgs-Boson searches as well as measurements in their own right. For this conference, several new measurements with the data from proton-proton collisions at $\sqrt{s} =$ 7 TeV taken in 2010 and 2011 with the ATLAS detector\cite{Aad:2008zzm} have been prepared: A measurement of the isolated di-photon cross-section\cite{CERN-PH-EP-2011-088} using 37 pb$^{-1}$ collected in 2010 and a measurement of the $W^\pm Z$ production cross-section\cite{ATLAS-CONF-2011-084} using 205 pb$^{-1}$ collected in 2011. In addition and for a complete view of the electroweak sector, the already published measurements of the $W\gamma$ and $Z\gamma$ cross-sections\cite{CERN-PH-EP-2011-079} and the $WW$ cross-sections\cite{CERN-PH-EP-2011-054} are presented as well.

\section{Di-photon cross-section}

Di-photon final states in proton-proton collisions may occur through quark-antiquark t-channel annihilation ($q\bar q \rightarrow \gamma\gamma$), or via gluon-gluon interactions ($gg \rightarrow \gamma\gamma$) by a quark box diagram. Even though the latter is of higher order, the high gluon flux at the LHC causes the two contributions to be comparable. 

\begin{figure}[tbhf]
\begin{center}
\epsfig{file=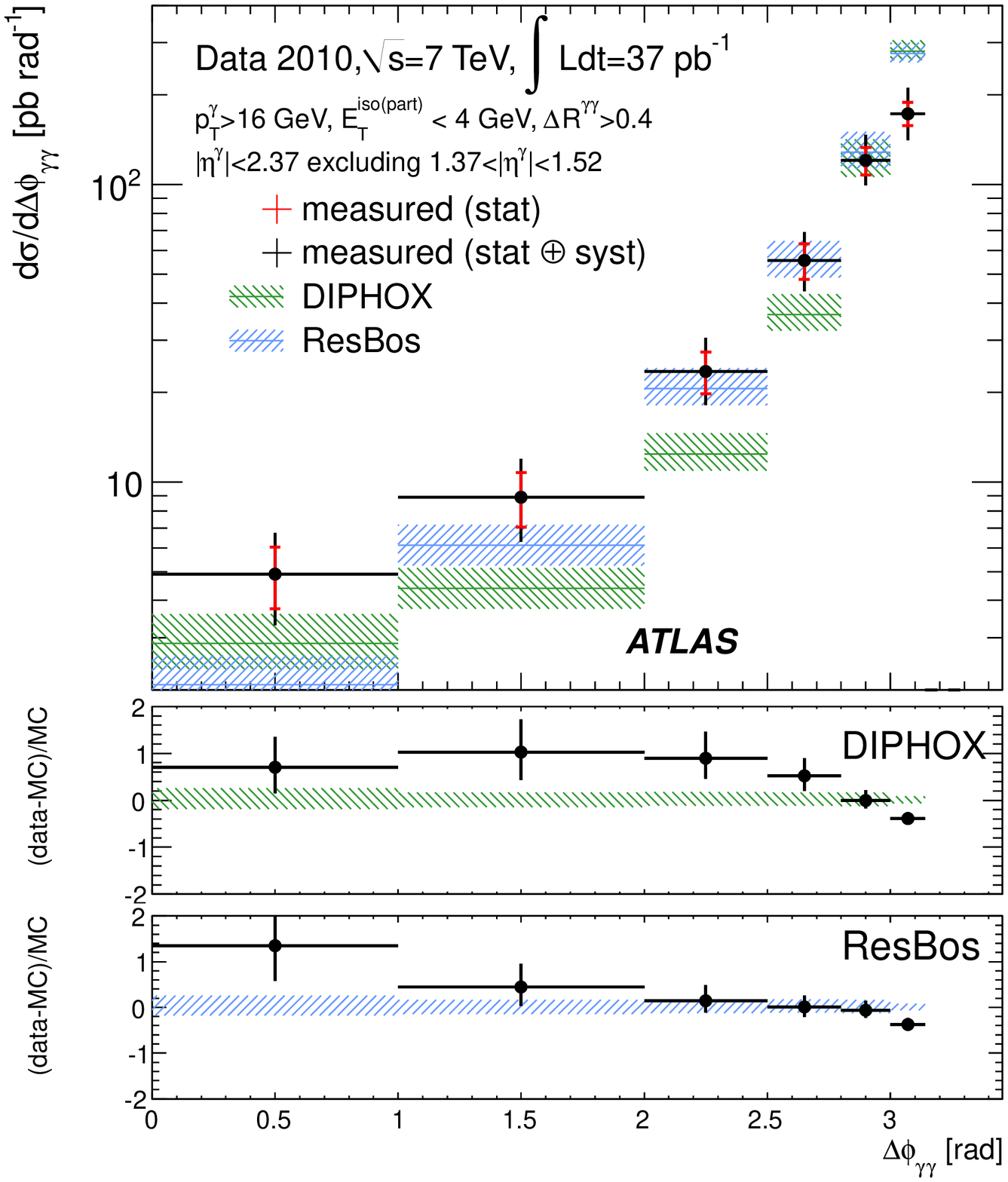,width=0.5\textwidth}\epsfig{file=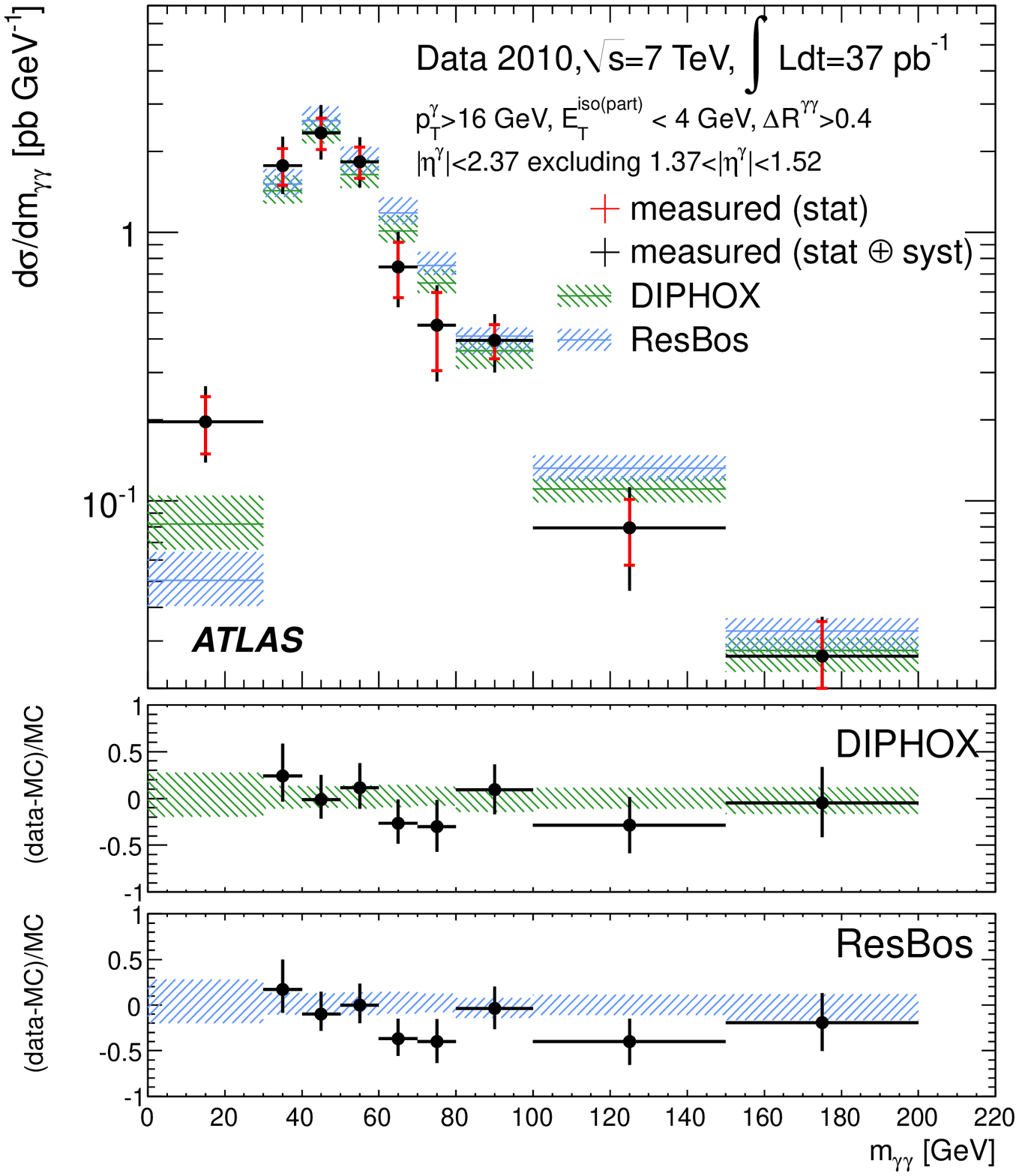,width=0.5\textwidth}
\caption{Differential di-photon cross-section  d$\sigma/$d$\Delta\varphi_{\gamma\gamma}$ and d$\sigma$/d$m_{\gamma\gamma}$}
\label{fig:ebkegamgam}
\end{center}
\end{figure}

In the selection of the photon pair candidates a fiducial acceptance region is required in pseudorapidity, $|\eta_\gamma| < 2.37$, excluding the barrel/endcap transition region $1.37 < |\eta_\gamma| < 1.52$. In addition, photon candidates are required to have a transverse momentum of more than 16 GeV. The opening angle of the photons must be $\Delta R_{\gamma\gamma} > 0.4$, and both photons have to be isolated: the transverse energy in a cone of angular radius $R < 0.4$ must be $E_T^{iso (part)} < 4$ GeV. The background, consisting of hadronic jets and isolated electrons, is estimated with fully data-driven techniques and subtracted. The selection of the control sample for this procedure is the main source of systematic uncertainty for this measurement.

Results of the measurement\cite{CERN-PH-EP-2011-088} are shown in Figure~\ref{fig:ebkegamgam} and compared with theoretical predictions from the DIPHOX and ResBos NLO generators. 
In the d$\sigma/$d$\Delta\varphi_{\gamma\gamma}$ differential cross-section the spectrum is broader towards low values of $\Delta\varphi_{\gamma\gamma}$ than NLO predictions, consistent with observations at the TeVatron\cite{Abazov:2010ah, Aaltonen:2011via}. This effect leads to the related discrepancy in the d$\sigma$/d$m_{\gamma\gamma}$ spectrum, which otherwise agrees well with predictions.

\section{$W^\pm Z$ cross-section}

The $W^\pm Z$ cross-section is measured using 205 pb$^{-1}$ of pp collision data taken in 2011\cite{CERN-PH-EP-2011-079}. In the analysis, three reconstructed leptons (e, $\mu$) are required, two of which must have the same flavor, opposite sign, and an invariant mass within 10 GeV of the Z mass. In addition, the missing tranverse energy $E_{T,miss}$ is required to be larger than 25\,GeV. In total, 12 candidates with an expected background of 2 events are observed.

The final result for the combined fiducial cross-section in this region is $\sigma^{fid}_{WZ\rightarrow \ell\nu\ell\ell} = 96^{+37}_{-30} \mathrm{(stat)} ^{+15}_{-14} \mathrm{(syst)} \pm 5 \mathrm{(lumi)}$ fb, with $\ell = e, \mu$, the dominant systematic uncertainty being the description of pile-up conditions for $E_{T,miss}$ with about 11\%. Extrapolating to the total cross-section gives $\sigma^{tot}_{WZ} = 18^{+7}_{-6} \mathrm{(stat)}^{+3}_{-3} \mathrm{(syst)} ^{+1}_{-1}\mathrm{(lumi)}$ pb, which agrees well with the Standard Model prediction of $16.9^{+1.2}_{-0.8}$ pb.

\section{$Z\gamma/W^\pm\gamma$ cross-section}
\begin{figure}[bt]
\begin{center}
\epsfig{file=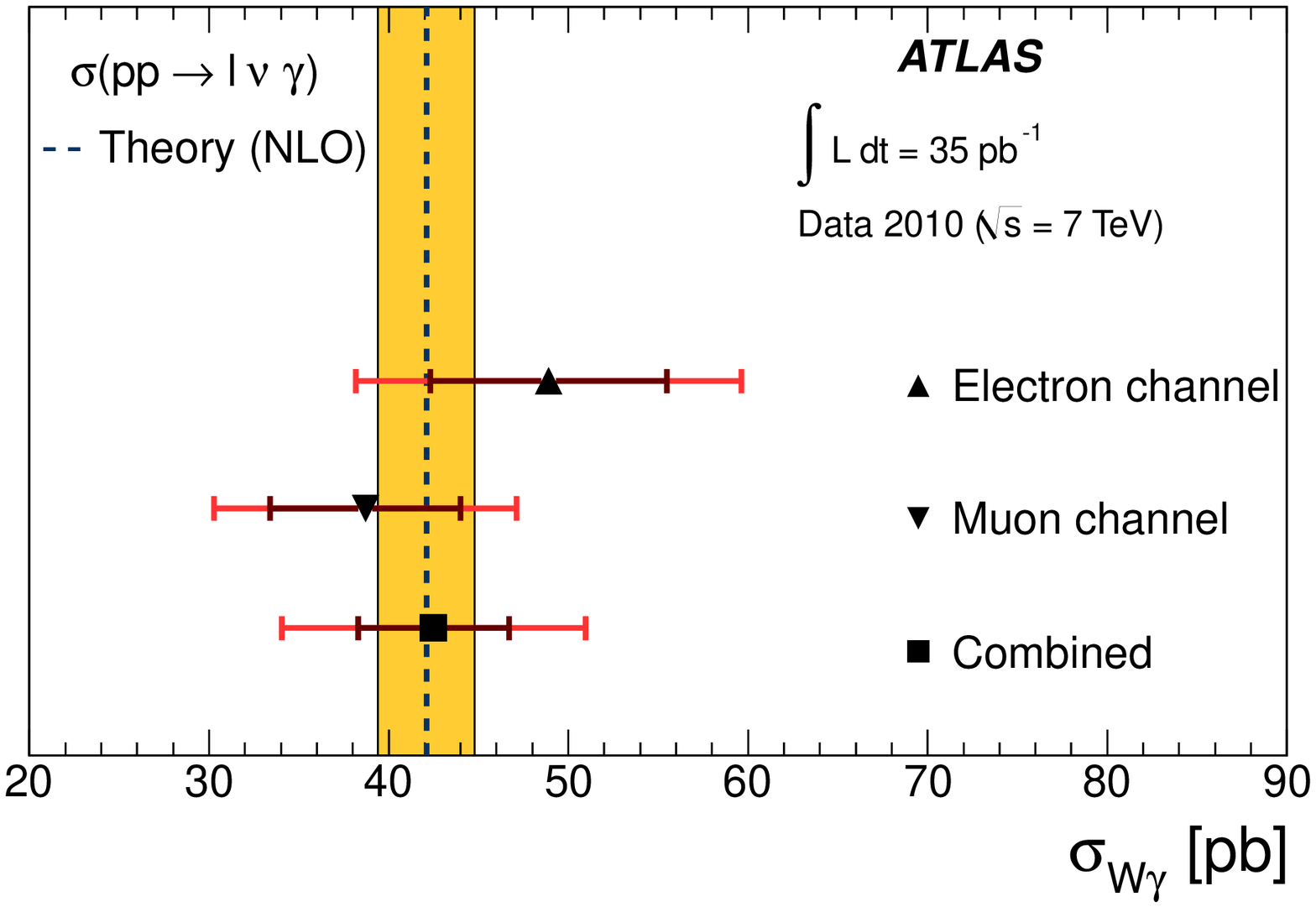,width=0.5\textwidth}\epsfig{file=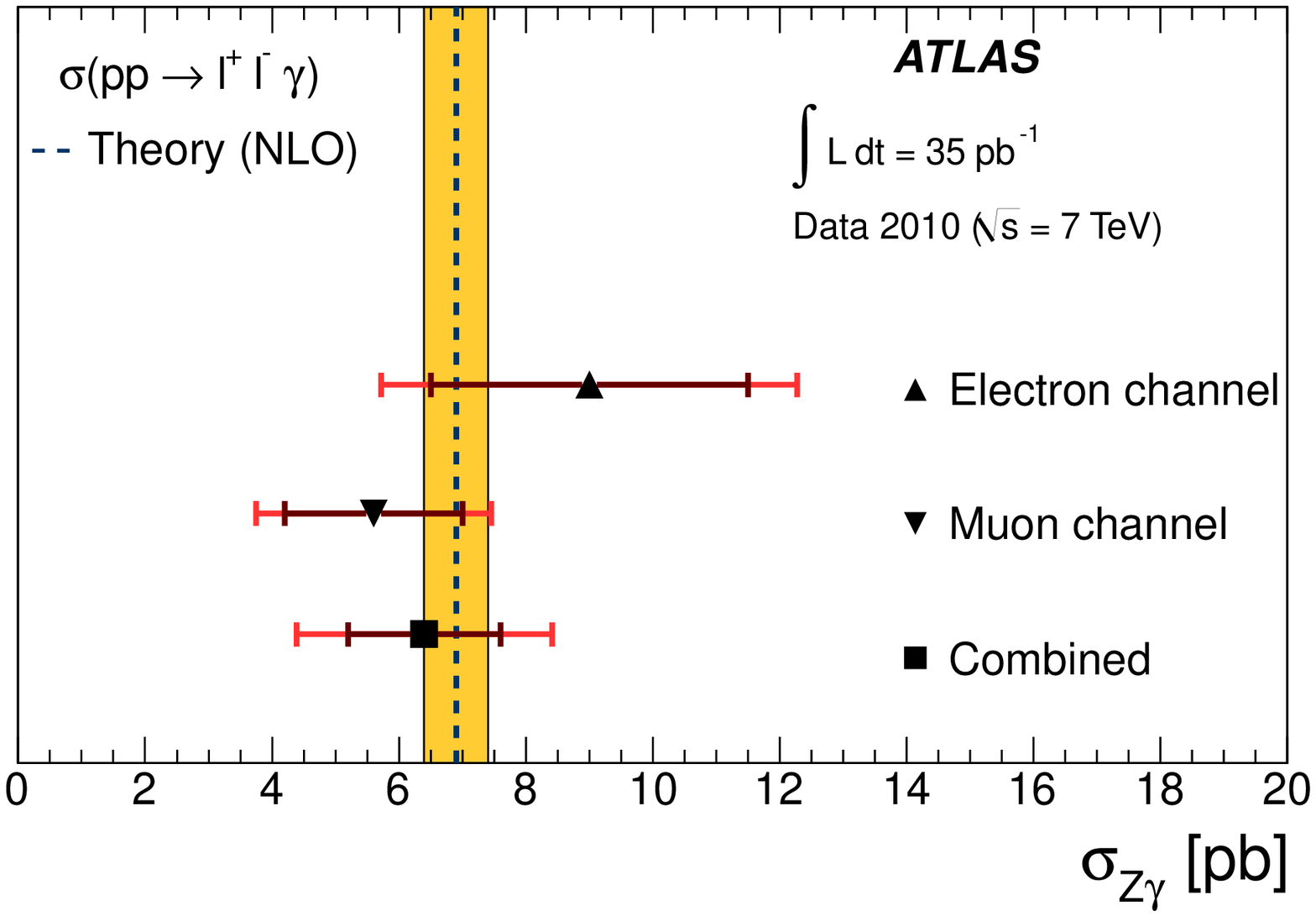,width=0.5\textwidth}
\caption{Summary of $Z^0\gamma/W^\pm\gamma$ cross-section measurements.}
\label{fig:ebkezgwg}
\end{center}
\end{figure}

Using $35$ pb$^{-1}$ of ATLAS data from 2010, the $Z\gamma/W^\pm\gamma$ cross-section measurement\cite{ATLAS-CONF-2011-084} closely follows the $W^\pm$ and $Z$ cross-section analyses, but requires one additional photon with a $p_T > 15$ GeV. To suppress the contribution of Final State Radiation (FSR) the $\Delta R$ between lepton and photon must be $\Delta R(\ell, \gamma) > 0.7$. The main backgrounds are estimated from data using a two-dimensional sideband method, being one of the main systematic uncertainties together with the uncertainty of photon identification efficiency. The measured cross-sections are compared to the SM predictions generated by NLO theory in Figure~\ref{fig:ebkezgwg} and agree well within uncertainties.

\section{$W^+W^-$ cross-section}

With $34$ pb$^{-1}$ of collision data from 2010 a first measurement at ATLAS of the $W^+W^-$ production cross-section is undertaken\cite{CERN-PH-EP-2011-054}. The analysis requires exactly two well-reconstructed oppositely charged leptons ($\ell$ = e, $\mu$) and excludes the Z resonance peak and low-mass resonances for leptons with same flavour. In addition, the component of the missing transverse energy perpendicular to the lepton closest in $\varphi$ must be larger than 20(40) GeV for leptons with different(same) flavour. Finally, all events with jets with $p_T > 20$ GeV are vetoed. The main systematic uncertainties stem from this jet veto (7.5\%) and from the lepton selection and identification (4\%). The background components are estimated using Monte Carlo and data driven methods for the $t\bar t$ and $W$+jets components. The cross-section is then determined by a maximum-likelihood fit combining the three channels. The resulting cross-section is $\sigma_{W^+W^-} = 41^{+20}_{-16}(stat)\pm5(syst)\pm1(lumi)$ pb, in good agreement with the QCD-NLO SM prediction of $44 \pm 3$ pb.

\section{Conclusion}

\begin{figure}[htb]
\begin{center}
\epsfig{file=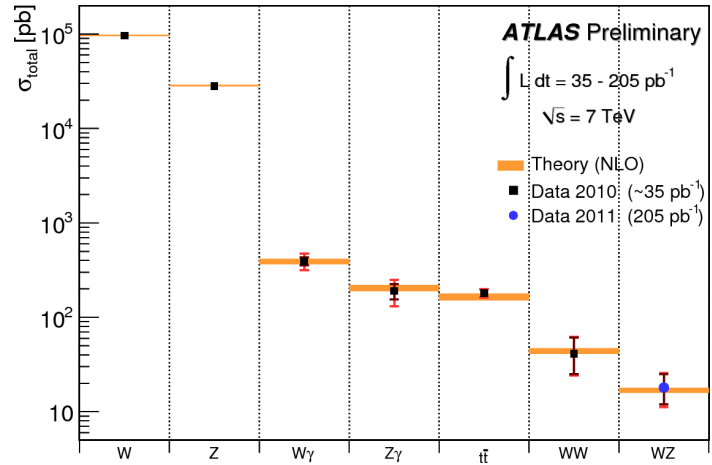,height=2.4in}
\caption{Summary of Electroweak measurements.}
\label{fig:ebkesummary}
\end{center}
\end{figure}

With collision data from the LHC taken in 2010 and 2011, we present measurements of $W\gamma$, $Z\gamma$, $W^+W^-$, $W^\pm Z$ and $\gamma\gamma$ production cross-sections. In Figure~\ref{fig:ebkesummary} these measurements are summarized and compared to NLO predictions. No deviations from Standard Model predictions are observed. For direct $\gamma\gamma$ production, differential cross-sections with respect to $m_{\gamma\gamma}$ and $\Delta\varphi_{\gamma\gamma}$ are shown. The next step will be updating all measurements using 2011 data, and setting limits on anomalous triple-gauge couplings.

%%%%%%%%%%%%%%%%%%%%%%%%%%%%%%%%%%%%%%%%%%%%%%%%%%%%%%%%%%%%%%%%%%%%%%%%%%%

\def\Discussion{
\setlength{\parskip}{0.3cm}\setlength{\parindent}{0.0cm}
     \bigskip\bigskip      {\Large {\bf Discussion}} \bigskip}
\def\speaker#1{{\bf #1:}\ }
\def\endDiscussion{}

%\Discussion
%\speaker{...}  ...

\endDiscussion
 
\end{document}